\documentclass{PoS}

\usepackage{lipsum}
\usepackage{lineno}

\usepackage{amsmath}
\usepackage{multicol}
\usepackage{mathtools}
\usepackage{cuted}
\usepackage{flushend}
\usepackage{amsmath}
\usepackage{graphicx}
\usepackage{bm}		
\usepackage{multirow}
\usepackage{enumitem}
\usepackage{graphicx}
\newcommand\varpm{\mathbin{\vcenter{\hbox{%
  \oalign{\hfil$\scriptstyle+$\hfil\cr
          \noalign{\kern-.3ex}
          $\scriptscriptstyle({-})$\cr}%
}}}}
\newcommand\varmp{\mathbin{\vcenter{\hbox{%
  \oalign{$\scriptstyle({+})$\cr
          \noalign{\kern-.3ex}
          \hfil$\scriptscriptstyle-$\hfil\cr}%
}}}}

\newcommand{\beq}{\begin{equation}}
\newcommand{\eeq}{\end{equation}}

\definecolor{draftcolor}{rgb}{0.5,0.5,0.8}

\title{Transverse momentum dependent multiplicities of hadrons produced in DIS at COMPASS}

\ShortTitle{Transverse momentum dependent multiplicities of hadrons produced in DIS at COMPASS}

\author{Andrea Moretti \\ 
        \rm{on behalf of the COMPASS Collaboration}\\
        \it{University and INFN, Trieste} \\
        E-mail: \email{andrea.moretti@ts.infn.it}}

\abstract{The COMPASS Collaboration measured the transverse momentum dependent multiplicities of charged hadrons produced in Deep Inelastic Scattering (DIS) off unpolarized protons. Complementing previous COMPASS measurements obtained with an isoscalar target, the data have been collected in 2016 and 2017 with 160 GeV/$c$ muon beams and a liquid hydrogen target. The multiplicities are studied as a function of the square of the hadron transverse momentum with respect to the virtual photon direction $P_{hT}^2$ in bins of the Bjorken variable $x$, of the photon virtuality $Q^2$ and of the fraction $z$ of the photon energy carried by the hadron. Preliminary results of this analysis, performed on a fraction of the available data sample, are shown here for the first time.  
}

\FullConference{XXVII International Workshop on Deep-Inelastic Scattering and Related Subjects - DIS2019\\
		8-12 April, 2019\\
		Torino, Italy}

\begin{document}

\section{Introduction}

The intrinsic motion of the parton in the nucleon is a fundamental ingredient in a complete description of the nucleon structure, which has to consider not only its projection along the nucleon momentum but also its transverse component in both momentum and coordinate space. The momentum distributions of the partons in the transverse plane is described by the Transverse-Momentum-Dependent Parton Distribution Functions (TMD-PDFs). Analogously, Transverse-Momentum-Dependent Fragmentation Functions (TMD-FFs) are necessary for a complete treatment of the hadron production. In this respects, the Semi-Inclusive Deep Inelastic Scattering (SIDIS) is a powerful tool to investigate both TMD-PDFs and TMD-FFs. In SIDIS events, a lepton scatters off a nucleon and at least one hadron is detected in the final state. Integrating over the azimuthal angle of the final state hadron and taking as nucleon an unpolarized proton, the differential cross section at twist-two reads \cite{Bacchetta:2006tn,Anselmino2014}:

\begin{equation}
    \frac{d^4\sigma^{h}} {dxdQ^2dzdP_{hT}^2} = \frac{2\pi^2\alpha^2}{(xs)^2}\frac{1+(1-y)^2}{y^2}F_{UU}(x,Q^2,z,P_{hT}^2)
\end{equation}
where $x$ is the Bjorken variable, $Q^2$ the photon virtuality, $z$ the fraction of the virtual photon energy carried by the hadron, $\textbf{P}_{hT}$ its transverse momentum with respect to the virtual photon, $y$ the fraction of the lepton energy carried by the photon and $s$ the energy in the center of mass squared, being $Q^2=xys$. The SIDIS structure function $F_{UU}$ can be written as:

\begin{align}
    F_{UU}(x,Q^2,z,P_{hT}^2) & = \sum_q e_q^2 \int d^2\textbf{k}_T d^2\textbf{p}_\perp \delta^{(2)}(\textbf{P}_{hT}-z\textbf{k}_T-\textbf{p}_\perp) f_1^q(x,Q^2,k_T)D_{1q}^h(x,Q^2,p_\perp) \\
    & = \sum_q e_q^2 f_1^q(x,Q^2,k_T) \otimes D_{1q}^h(x,Q^2,p_\perp) 
\end{align}
where the convolution integral is performed over the unobservable transverse momentum of the quark $\textbf{k}_T$ and over the transverse momentum acquired in the fragmentation process $\textbf{p}_\perp$, related by $\textbf{P}_{hT}=z\textbf{k}_T+\textbf{p}_\perp$; $f_1^q(x,Q^2,k_T)$ is the unpolarized TMD-PDF and $D_{1q}^h(x,Q^2,p_\perp)$ the unpolarized TMD-FF. The hadron multiplicities $\mathcal{M}^h$ are defined as the ratio of the SIDIS cross section over the DIS cross section:

\begin{equation}
    \frac{d^2\mathcal{M}^h(x,Q^2,z,P_{hT}^2)}{dzdP_{hT}^2} = \left(\frac{d^4\sigma^h}{dxdQ^2dzdP_{hT}^2}\right) \Bigg{/} \left(\frac{d^2\sigma^{DIS}}{dxdQ^2}\right)  = \frac{\sum_q e_q^2 f_1^q(x,Q^2,k_T) \otimes D_{1q}^h(x,Q^2,p_\perp)}{\sum_q e_q^2 f_1^q(x,Q^2)}, \\
\end{equation}
where $f_1^q(x,Q^2)$ is the unpolarized transverse momentum independent PDF. The multiplicities give thus access to the transverse momenta which, by the way, can not be disentangled with such measurement in a model independent way. In the ratio of the two cross sections many common terms (like the luminosity and the efficiencies in the muon detection, trigger and DAQ) cancel out and in each $x$, $Q^2$, $z$ and $P_{hT}^2$ bin the multiplicities can be measured as:

\begin{equation}
    \frac{d^2\mathcal{M}^h(x,Q^2,z,P_{hT}^2)}{dzdP_{hT}^2} = \frac{N^h(x,Q^2,z,P_{hT}^2)}{N^{DIS}(x,Q^2)\Delta z \Delta P_{hT}^2} \frac{1}{a^h(x,Q^2,z,P_{hT}^2)}, 
\end{equation}
being $N^h$ the number of charged hadrons and $N^{DIS}$ the number of DIS events, $\Delta z$ and $\Delta P_{hT}^2$ the bin widths and $a^h$ the hadron acceptance.

The expression for the charged hadrons multiplicities deserves further corrections, due to the radiative effects, to the electron contamination and to the diffractively produced vector mesons \cite{Kerbizi:2018bvk}. Such corrections are at the moment estimated but not applied, in view of more detailed studies to come. Here we present preliminary results obtained  in a kinematic region where such corrections are estimated to be small. The results are obtained from about 10\% of the data collected at COMPASS during the 2016 run with a 160 GeV/$c$ muon beam (both $\mu^+$ and $\mu^-$ beams with balanced statistics) and a 2.5 m long liquid hydrogen target.

\section{Extraction of hadron multiplicities}
\subsection{Event and hadron selection}
 
\begin{figure}[!h]
\begin{center}
      \includegraphics[width=0.35\textwidth]{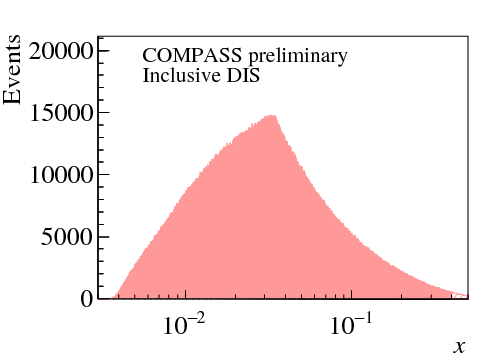} \includegraphics[width=0.35\textwidth]{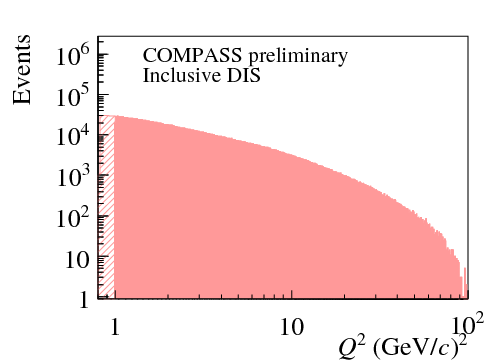}\\
    \includegraphics[width=0.35\textwidth]{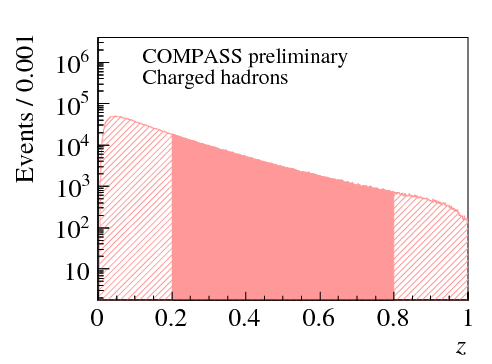}
    \includegraphics[width=0.35\textwidth]{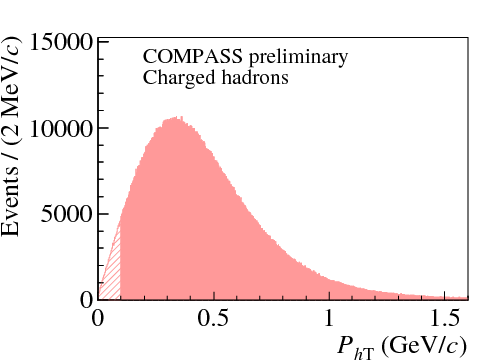}
    \caption{Top row: the kinematic distributions of $x$ and $Q^2$ of the selected DIS events. Bottom row: the $z$ and $P_{hT}$ distributions of the selected charged hadrons. The shaded regions are removed with proper cuts.}
    \label{fig:kin_dist}
\end{center}{}
  \end{figure}

The selection of DIS events has been performed asking for $Q^2>1$ (GeV/$c)^2$, $0.1<y<0.9$, mass of the hadronic final state $W>3$ GeV/$c^2$ and $0.003<x<0.4$. Hadrons have been selected by requiring the material integrated over the path of the reconstructed trajectories not to exceed 10 radiation lengths and with $z$ and $P_{hT}$ in the ranges: $0.2<z<0.8$ and $P_{hT}>0.1$ GeV/$c$. Plots of kinematic distributions of $x$, $Q^2$, $z$ and $P_{hT}$ are given in Fig. \ref{fig:kin_dist}. The kinematic ranges shown in full color in Fig. \ref{fig:kin_dist} correspond to the aforementioned kinematic cuts.

\subsection{Acceptance, vector meson contamination and choice of the range}

The hadron acceptance term $a^h$ has been estimated with a Monte Carlo simulation based on the LEPTO generator \cite{Ingelman:1996mq} and defined as the ratio of reconstructed and generated hadrons, provided the corresponding DIS event was reconstructed:

\begin{equation}
    a^h(x,Q^2,z,P_{hT}^2) = \frac{N^h_{rec}(x,Q^2,z,P_{hT}^2)}{N^h_{gen}(x,Q^2,z,P_{hT}^2)|_{DIS_{rec}}}.
\end{equation}

The acceptance shows a flat trend in $P_{hT}^2$ in the ($x,Q^2$) range $0.003<x<0.008$ and $16<Q^2/(\mathrm{GeV}/c)^2<81$ and for all hadrons with $z>0.3$. Due to the 2016/2017 reduced geometrical acceptance of the spectrometer, this is not the case for hadrons with low $y$ and simultaneously $0.2<z<0.3$, which are then not considered in this work. Further selections on the DIS and hadron samples (like on the position of the vertex in the target) are expected to increase the acceptance in future analyses and are currently under study.

It is at high $z$, on the other hand, that the fraction of hadrons originating from the decay of vector mesons diffractively produced shows its maximum. This is mostly due to the decay of $\rho$ meson into a $\pi^+\pi^-$ pair, while a limited contamination at intermediate $z$ is induced by $K^+K^-$ pairs produced in the decay of a $\phi$ meson. The contamination has been estimated combining the predictions from two HEPGEN Monte Carlo samples \cite{Sandacz:2012at} (one for $\rho$ and one for $\phi$ mesons) and a LEPTO sample. Since the contamination increases with $z$, these preliminary hadron multiplicities are extracted up to $z=0.6$, i.e. in a range where it is estimated to be always smaller than 5\%.

\subsection{Results}
\begin{figure}[!h]
\begin{center}
      \includegraphics[width=0.75\textwidth]{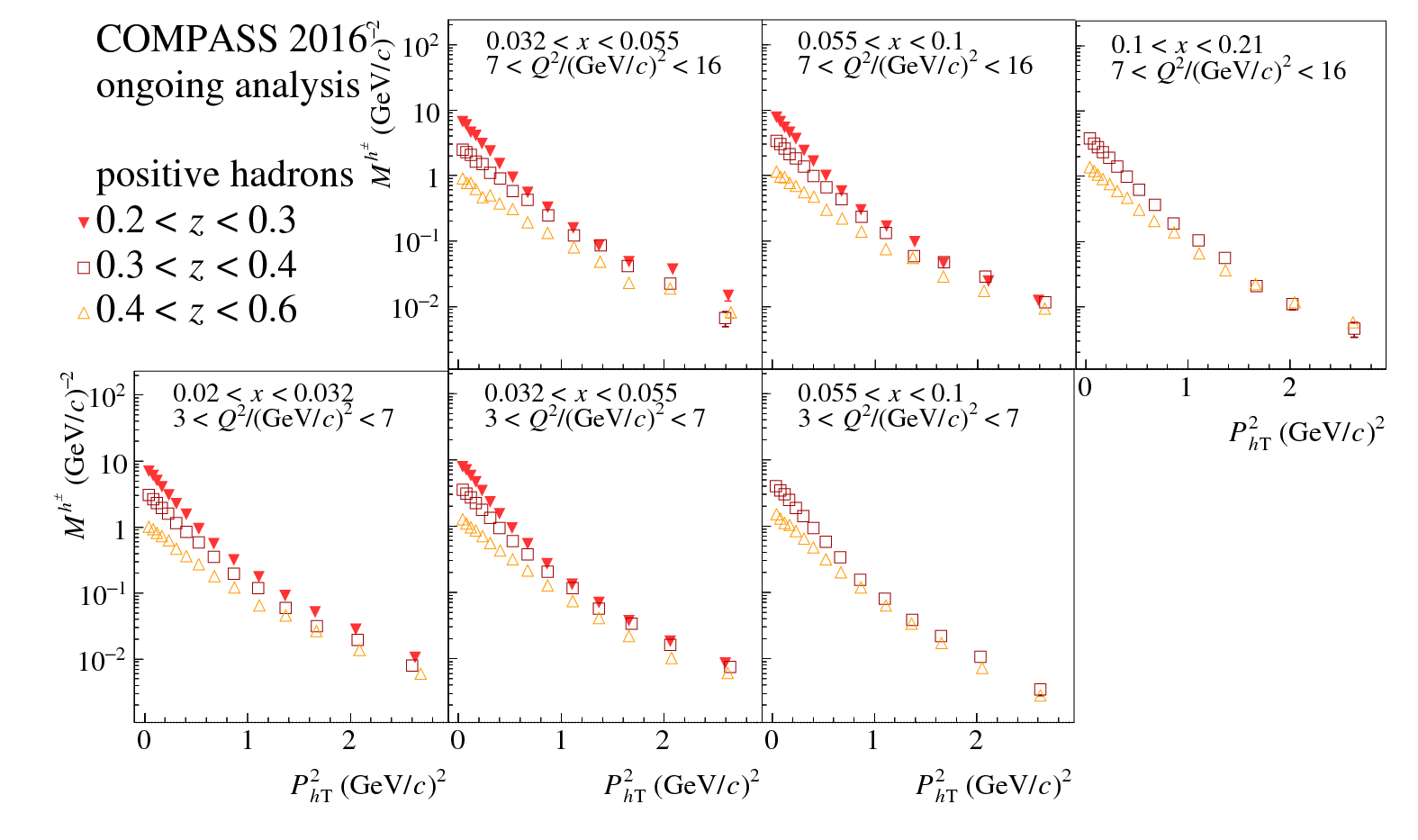}
     \caption{Unidentified positively-charged hadron multiplicities as a function of $P_{hT}^2$ in selected bins of $x$, $Q^2$ and $z$. Not corrected for radiative effects.}
    \label{fig:multp}  
\end{center}{}
\end{figure}

\begin{figure}[!h]
\begin{center}
      \includegraphics[width=0.75\textwidth]{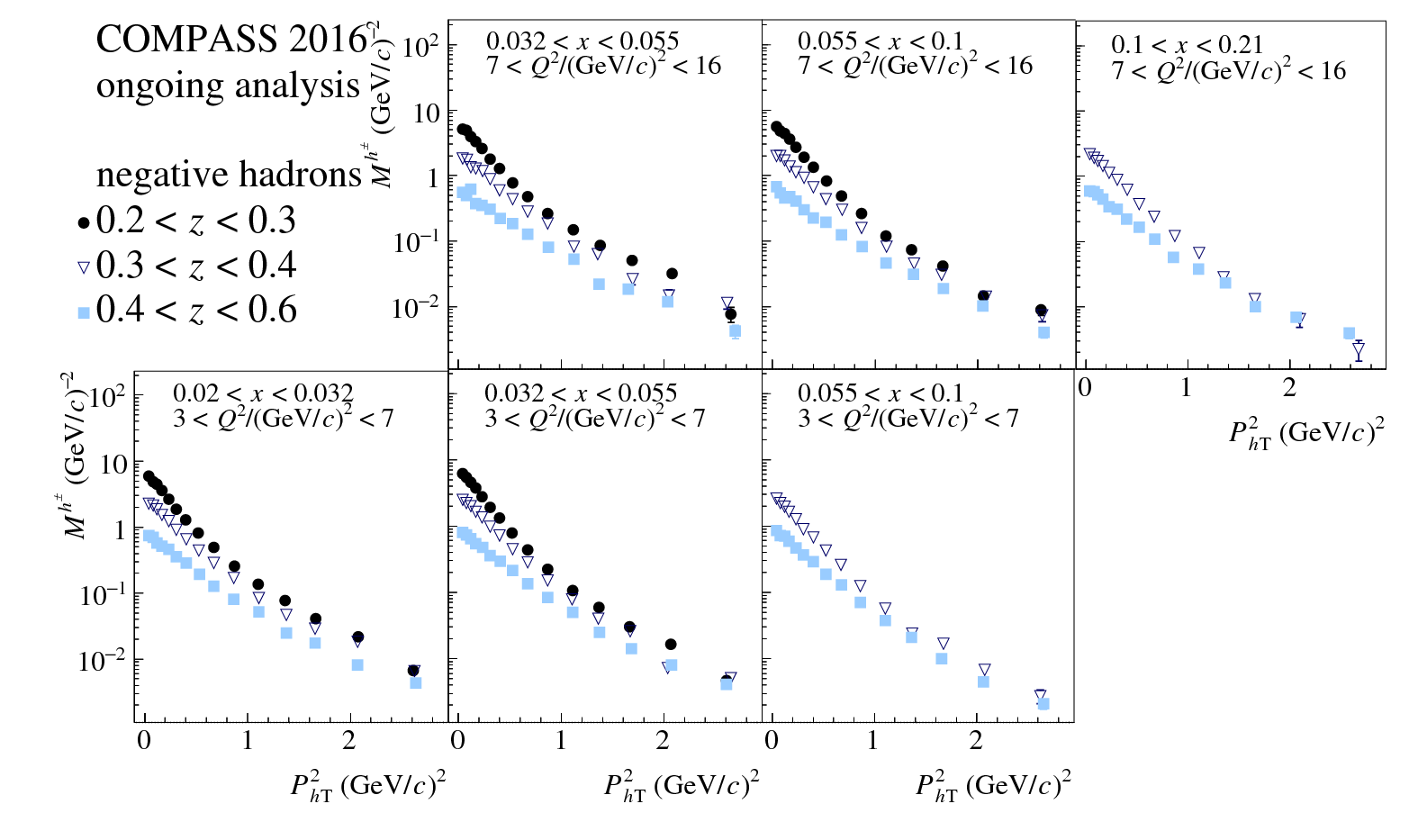} 
    \caption{Unidentified negatively-charged hadron multiplicities as a function of $P_{hT}^2$ in selected bins of $x$, $Q^2$ and $z$. Not corrected for radiative effects.}
    \label{fig:multm}  
\end{center}{}
\end{figure}

The multiplicities of positive and negative hadrons in the four-dimensional grid of kinematic bins are shown in Figs. \ref{fig:multp} and \ref{fig:multm}. For each 3D bin in $x, Q^2$ and $z$ the multiplicities are shown as a function of $P_{hT}^2$; different $z$ bins are superimposed in the same pads to better show the evolution with $z$. Uncertainties are statistical only; the systematic uncertainty is estimated to be smaller than 10\% of the multiplicity value. Current results are not corrected for radiative effects.

\begin{figure}[!tbp]
  \centering
  \begin{minipage}[b]{0.49\textwidth}
   \includegraphics[width=\textwidth]{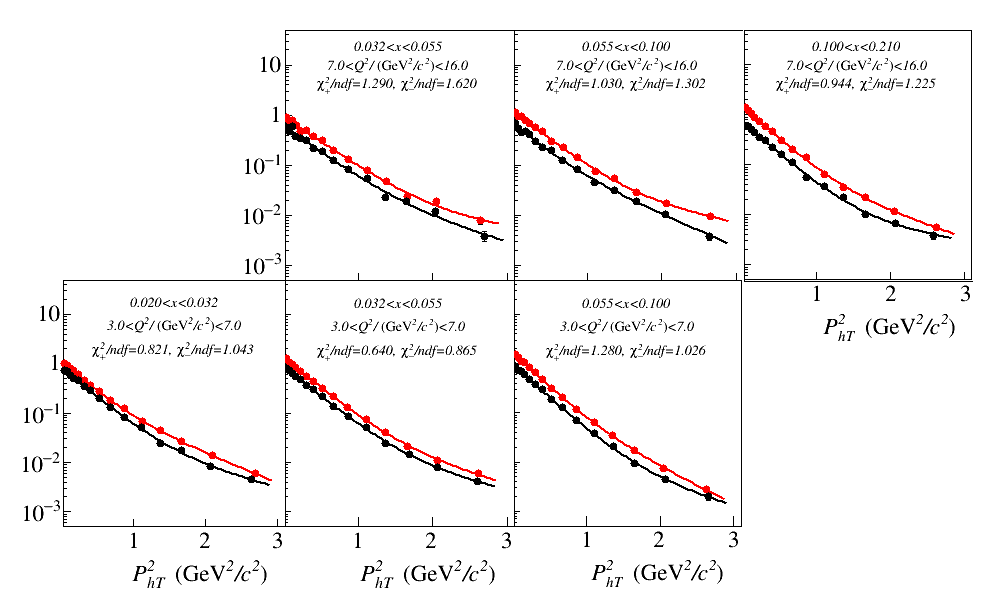}
 \caption{Fit of the hadron multiplicities for $0.4<z<0.6$ with the double exponential function $f_1$ for positive (red) and negative hadrons (black).
\label{fig:fit1}}
  \end{minipage}
  \hfill
  \begin{minipage}[b]{0.49\textwidth}
    \includegraphics[width=\textwidth]{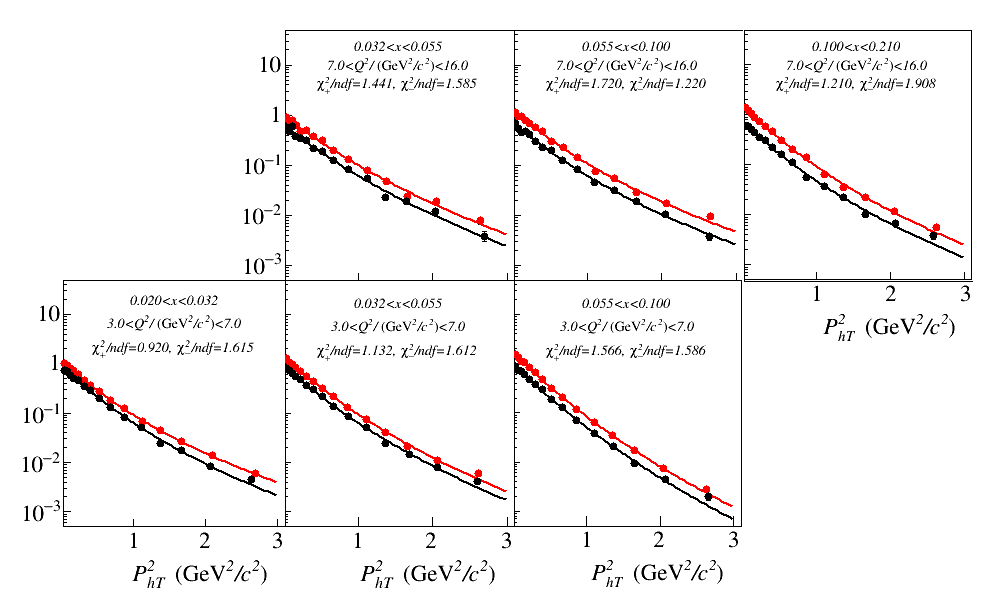}
 \caption{Fit of the hadron multiplicities for $0.4<z<0.6$ with the Tsallis function $f_2$  for positive (red) and negative hadrons (black).
\label{fig:fit2}}
  \end{minipage}
\end{figure}

The hadron multiplicities have been fitted with two different functions, namely with a "double exponential" function $f_1$,  and with the Tsallis function $f_2$:


\begin{equation}
    f_1 = \frac{N_1}{\alpha_1}\exp\left(-\frac{P_{hT}^2}{\alpha_1}\right) +\frac{N_1^\prime}{\alpha_1^\prime} \exp\left(-\frac{P_{hT}^2}{\alpha_1^\prime}\right), \hspace{0.5cm}
\end{equation}

\begin{equation}
    f_2 = N_2\left(1-(1-q)\frac{P_{hT}^2}{T}\right)^{\frac{1}{1-q}}.
\end{equation}
Both functions aim at describing the two different kinds of power-law behaviour observed for small and larger $P_{hT}$. At variance with $f_1$, $f_2$ encodes the behaviour in the two regions in a single functional form. In the corresponding bin, the values of the fitted parameters are in agreement with the published ones \cite{Aghasyan:2017ctw}. The fits of the multiplicities in the third $z$ bin ($0.4<z<0.6$) are shown in Figs. \ref{fig:fit1} and \ref{fig:fit2} for both functions.

\newpage

\section{Conclusions and perspectives}

Transverse momentum dependent hadron multiplicities are fundamental tools towards a thorough understanding of the nucleon structure: as such, they have aroused a growing interest in recent years both on the theoretical and on the experimental side. The COMPASS Collaboration contributed to this topic with a measurement of the multiplicities on an isoscalar target and it is still actively working on the subject. We presented here for the first time COMPASS preliminary results on the multiplicities of charged hadrons produced with muon beams on an unpolarized proton target. The multiplicities have been extracted as a function of the squared hadron transverse momentum $P_{hT}^2$ in three dimensional bins of $x$, $Q^2$ and $z$. The current analysis has been performed on part of the data collected in 2016, in a selected kinematic range where acceptance is high and almost flat in $P_{hT}^2$ and where possible contaminations have been estimated to be small. This analysis will be extended to a wider kinematic range and to the full 2016 and 2017 data set. The hadron azimuthal asymmetries, already measured by COMPASS on an iscoscalar target \cite{Adolph:2014pwc} and still currently studied \cite{Moretti:2019lkw,Jan:2019dis}, will be measured from the same proton data in the same $x$, $Q^2$ and $z$ bins. All these new results, together with other results by COMPASS, HERMES and JLab, will allow for a complete analysis and possibly lead to an extraction of the still unknown Boer-Mulders TMD. 
\bigskip


\begin{thebibliography}{1}

\bibitem{Bacchetta:2006tn}
A.~Bacchetta, M.~Diehl, K.~Goeke, A.~Metz, P.~J. Mulders and M.~Schlegel,
  \emph{{Semi-inclusive deep inelastic scattering at small transverse
  momentum}}, \href{https://doi.org/10.1088/1126-6708/2007/02/093}{\emph{JHEP}
  {\bfseries 02} (2007) 093}
  [\href{https://arxiv.org/abs/hep-ph/0611265}{{\ttfamily hep-ph/0611265}}].

\bibitem{Anselmino2014}
M.~Anselmino, M.~Boglione, J.~O. Gonzalez~H., S.~Melis and A.~Prokudin,
  \emph{Unpolarised transverse momentum dependent distribution and
  fragmentation functions from sidis multiplicities},
  \href{https://doi.org/10.1007/JHEP04(2014)005}{\emph{JHEP} {\bfseries 2014}
  (2014) 5}.

\bibitem{Kerbizi:2018bvk}
A.~Kerbizi~on~behalf~of~the COMPASS~Collaboration, \emph{{Interpretation of the
  unpolarized azimuthal asymmetries in SIDIS}},  in \emph{{23rd International
  Symposium on Spin Physics (SPIN 2018) Ferrara, Italy, September 10-14,
  2018}}, 2018, \href{https://arxiv.org/abs/1812.07477}{{\ttfamily
  1812.07477}}.

\bibitem{Ingelman:1996mq}
G.~Ingelman, A.~Edin and J.~Rathsman, \emph{{LEPTO 6.5: A Monte Carlo generator
  for deep inelastic lepton - nucleon scattering}},
  \href{https://doi.org/10.1016/S0010-4655(96)00157-9}{\emph{Comput. Phys.
  Commun.} {\bfseries 101} (1997) 108}
  [\href{https://arxiv.org/abs/hep-ph/9605286}{{\ttfamily hep-ph/9605286}}].

\bibitem{Sandacz:2012at}
A.~Sandacz and P.~Sznajder, \emph{{HEPGEN - generator for hard exclusive
  leptoproduction}},  \href{https://arxiv.org/abs/1207.0333}{{\ttfamily
  1207.0333}}.

\bibitem{Aghasyan:2017ctw}
M.~Aghasyan~et~al. (COMPASS~Collaboration),
  \emph{{Transverse-momentum-dependent Multiplicities of Charged Hadrons in
  Muon-Deuteron Deep Inelastic Scattering}},
  \href{https://doi.org/10.1103/PhysRevD.97.032006}{\emph{Phys. Rev.}
  {\bfseries D97} (2018) 032006}
  [\href{https://arxiv.org/abs/1709.07374}{{\ttfamily 1709.07374}}].

\bibitem{Adolph:2014pwc}
C.~Adolph~et~al. (COMPASS~Collaboration), \emph{{Measurement of azimuthal
  hadron asymmetries in semi-inclusive deep inelastic scattering off
  unpolarised nucleons}},
  \href{https://doi.org/10.1016/j.nuclphysb.2014.07.019}{\emph{Nucl. Phys.}
  {\bfseries B886} (2014) 1046}
  [\href{https://arxiv.org/abs/1401.6284}{{\ttfamily 1401.6284}}].

\bibitem{Moretti:2019lkw}
A.~Moretti~on~behalf~of~the COMPASS~Collaboration, \emph{{Measurement of
  azimuthal asymmetries in SIDIS on unpolarized protons}},  in \emph{{23rd
  International Symposium on Spin Physics (SPIN 2018) Ferrara, Italy, September
  10-14, 2018}}, 2019, \href{https://arxiv.org/abs/1901.01773}{{\ttfamily
  1901.01773}}.

\bibitem{Jan:2019dis}
J.~Matousek~on~behalf~of~the COMPASS~Collaboration, \emph{{Measurement of the
  azimuthal modulations of hadrons in unpolarised SIDIS}}, {\emph{these
  Proceedings} }.

\end{thebibliography}
\bibliographystyle{jhep.bst}

\providecommand{\href}[2]{#2}\begingroup\raggedright\endgroup

\end{document}